\documentclass[useAMS,usenatbib]{mn2e}
\usepackage{psfrag,graphicx}
\usepackage{color}
\usepackage{txfonts}
\usepackage{breqn}
\usepackage[T1]{fontenc}
\usepackage{ae,aecompl}
\usepackage{graphicx}
\title[Black hole formation in the early universe]
   {Impact of baryonic streaming velocities on the formation of supermassive black holes via direct collapse}
\author[Latif et al.]
  {M.~A.~Latif,$^1$
  J.~C.~Niemeyer,$^1$
  D.~R.~G.~Schleicher$^1$ 
     \newauthor 
   $^1$ Institut f\"ur Astrophysik, Georg-August-Universit\"at, \\
    Friedrich-Hund-Platz 1, D-37077 G\"ottingen, Germany}
\date{}

\pagerange{\pageref{firstpage}--\pageref{lastpage}} \pubyear{2009}

\def\LaTeX{L\kern-.36em\raise.3ex\hbox{a}\kern-.15em
      T\kern-.1667em\lower.7ex\hbox{E}\kern-.125emX}

\begin{document}

\bibliographystyle{mn2e}

\label{firstpage}

\maketitle
\def\na{NewA}%
\def\aj{AJ}%
\def\araa{ARA\&A}%
\def\apj{ApJ}%
\def\apjl{ApJ}%
\def\jcap{JCAP}

\def\apjs{ApJS}%
\def\ao{Appl.~Opt.}%
\def\apss{Ap\&SS}%
\def\aap{A\&A}%
\def\aapr{A\&A~Rev.}%
\def\aaps{A\&AS}%
\def\azh{AZh}%
\def\baas{BAAS}%
\def\jrasc{JRASC}%
\def\memras{MmRAS}%
\def\mnras{MNRAS}%
\def\pra{Phys.~Rev.~A}%
\def\prb{Phys.~Rev.~B}%
\def\prc{Phys.~Rev.~C}%
\def\prd{Phys.~Rev.~D}%
\def\pre{Phys.~Rev.~E}%
\def\prl{Phys.~Rev.~Lett.}%
\def\pasp{PASP}%
\def\pasj{PASJ}%
\def\qjras{QJRAS}%
\def\skytel{S\&T}%
\def\solphys{Sol.~Phys.}%

\def\sovast{Soviet~Ast.}%
\def\ssr{Space~Sci.~Rev.}%
\def\zap{ZAp}%
\def\nat{Nature}%
\def\iaucirc{IAU~Circ.}%
\def\aplett{Astrophys.~Lett.}%
\def\apspr{Astrophys.~Space~Phys.~Res.}%
\def\bain{Bull.~Astron.~Inst.~Netherlands}%
\def\fcp{Fund.~Cosmic~Phys.}%
\def\gca{Geochim.~Cosmochim.~Acta}%
\def\grl{Geophys.~Res.~Lett.}%
\def\jcp{J.~Chem.~Phys.}%
\def\jgr{J.~Geophys.~Res.}%
\def\jqsrt{J.~Quant.~Spec.~Radiat.~Transf.}%
\def\memsai{Mem.~Soc.~Astron.~Italiana}%
\def\nphysa{Nucl.~Phys.~A}%
\def\physrep{Phys.~Rep.}%
\def\physscr{Phys.~Scr}%
\def\planss{Planet.~Space~Sci.}%
\def\procspie{Proc.~SPIE}%

%


\begin{abstract}
{Baryonic streaming motions produced prior to the epoch of recombination became supersonic during the cosmic dark ages. Various studies suggest that such streaming velocities change the halo statistics and also influence the formation of Population III stars. In this study, we aim to explore the impact of streaming velocities on the formation of supermassive black holes at $z>10$ via the direct collapse scenario. To accomplish this goal, we perform cosmological large eddy simulations for two halos of a few times $\rm 10^{7}~M_{\odot}$ with initial streaming velocities of 3, 6 and 9 $\rm km/s$. These massive primordial halos illuminated by the strong Lyman Werner flux are the potential cradles for the formation of direct collapse seed black holes. To study the evolution for longer times, we employ sink particles and track the accretion for 10,000 years. Our findings show that higher streaming velocities increase the circular velocities from about 14 $\rm km/s$ to 16 $\rm km/s$. They also delay the collapse of halos for a few million years, but do not have any significant impact on the halo properties such as turbulent energy, radial velocity, density and accretion rates. Sink particles of about $\rm \sim 10^5~M_{\odot}$ are formed at the end of our simulations and no clear distribution of sink masses is observed in the presence of streaming motions. It is further found that the impact of streaming velocities is less severe in massive halos compared to the minihalos as reported in the previous studies.  

}
\end{abstract}



\section{Introduction}
Baryonic acoustic oscillations prior to the epoch of recombination generated baryonic streaming motions of $\rm \sim 30~ km/s$ \citep{2010PhRvD..82h3520T}. During recombination, the radiation dominated plasma  transformed to neutral gas and the sound speed dropped from relativistic to thermal velocities of $2 \times 10^{-5}c$. The latter is smaller than the relative motion of gas and dark matter particles. Consequently, the streaming motions of baryons became supersonic with typical Mach numbers of about 5. Such streaming velocities impede small scale density perturbations,  may lead to the suppression of halo abundances and also enhance the bias and clustering of the halos \citep{2010PhRvD..82h3520T,2010JCAP...11..007D,2011MNRAS.412L..40M,2012ApJ...760....4O,2012MNRAS.424.1335F}. This bulk motion may further move the baryons out of the dark matter potentials. Their impact on the formation of supermassive black holes in massive primordial halos remains unexplored. 

Population III stars are the first sources of light to be formed at the end of cosmic dark ages \citep{2000ApJ...540...39A,Clark11,Greif12,2012MNRAS.422..290S,2013MNRAS.434L..36B,2013ApJ...772L...3L,2013arXiv1307.7567B}. The role of streaming velocities in the context of Population III star formation in minihalos of $\rm 10^5-10^6~M_{\odot}$ has been studied by \cite{2011ApJ...730L...1S} and \cite{2011ApJ...736..147G}, and they found that typical streaming velocities of $\rm3~km/s$ at $z=100$ delay the formation of the first stars, and increase the virial mass required for molecular hydrogen cooling to become effective. The enhanced mass for $\rm H_{2}$ cooling  may influence the statistics of minihalos and even the presence of higher turbulent energy may reduce the masses of the first stars by inducing fragmentation. In a recent study, \cite{2013MNRAS.435.3559T} have found that streaming motions significantly reduce the number density of stellar seed black hole at $z> 15$, while their overall impact on the formation of high-redshift BHs is negligible. They have also proposed that streaming velocities of 2-3 times the root mean square value could delay the formation of stars in minihalos until the halo mass reaches the threshold value for atomic cooling and may further facilitate the formation of supermassive black holes by the direct collapse \citep{2013arXiv1310.0859T}. It is expected that streaming motions may enhance the turbulent accretion and lead to higher seed black hole masses.

The existence of $\rm \sim 10^{9}~M_{\odot}$ supermassive black holes has been revealed from the observations of quasars at $z>6$ \citep{2003AJ....125.1649F,2006AJ....131.1203F,2010AJ....139..906W,2011Natur.474..616M,2013arXiv1311.3666V}. How such massive objects are assembled in the infant Universe and what their potential progenitors are remains an unfathomable conundrum. Numerous theoretical models propose various pathways such as accretion and merging of stellar mass black holes \citep{2001ApJ...552..459H,2004ApJ...613...36H,2009ApJ...696.1798T,2012ApJ...756L..19W}, run-away collapse of dense stellar cluster due to the relativistic instability \citep{2004Natur.428..724P,2008ApJ...686..801O, 2009ApJ...694..302D} and the direct collapse of a protogalactic gas cloud \citep{2002ApJ...569..558O,2003ApJ...596...34B,2006ApJ...652..902S,2006MNRAS.370..289B,2006MNRAS.371.1813L,2008MNRAS.391.1961D,2008arXiv0803.2862D,2010MNRAS.402.1249S,2010MNRAS.tmp.1427J,2010ApJ...712L..69S,2011MNRAS.411.1659L,2013arXiv1301.5567P,2013MNRAS.433.1607L,2013MNRAS.tmp.2526L,2013ApJ...774...64W,2013ApJ...771...50A,2013arXiv1309.1067S,2013arXiv1310.3680L}. The growth of stellar mass black holes is extremely challenging as they have to accrete at the Eddington limit almost all the time to reach the observed masses. On the other hand, the direct collapse model provides massive seeds of $\rm 10^4-10^6~M_{\odot}$ which may grow at relatively moderate accretion rates to form billion solar mass black holes.

Massive primordial halos of $\rm 10^{7}-10^{8}~M_{\odot}$ formed at $z=15$ and irradiated by the strong Lyman-Werner flux are the potential cradles for the birth of supermassive black holes forming via the direct collapse scenario. Such conditions can be achieved in the early Universe and their occurrence is frequent enough to produce the observed number density of black holes \citep{2008MNRAS.391.1961D,2010MNRAS.402.1249S,2012MNRAS.425.2854A}, see also \cite{2012MNRAS.422.2539I,2013A&A...553L...9V}. Numerical simulations studying the collapse of such halos show that fragmentation remains suppressed in the presence of a strong photo-dissociating background flux and massive objects are likely to be formed \citep{2003ApJ...596...34B,2008ApJ...682..745W,2009MNRAS.393..858R,2011MNRAS.411.1659L,2013MNRAS.433.1607L}. In a recent study,  \cite{2013MNRAS.tmp.2526L} have evolved the simulations for $\rm 2 \times 10^{4}$ years after the initial collapse and have shown that seed black holes of about $\rm 10^{5}~M_{\odot}$ are formed. This study further demonstrates the feasibility of the direct collapse scenario.

In this article, we explore the impact of baryonic streaming velocities on the formation of seed black holes via the direct collapse mechanism. To achieve this goal, we perform large eddy simulations (LES) with initial streaming velocities of $\rm 30~\times(z/1000)~km/s$. To further investigate the role of large streaming velocities, we perform comparison runs with streaming velocities of  $\rm 60 ~\times(z/1000)~km/s$ and $\rm 90~\times(z/1000)~km/s$. We make use of sink particles to follow the accretion for longer times and employ a fixed Jeans resolution of 32 cells during the entire course of the simulations. This study allows us to investigate the role of baryonic streaming motions in the assembling of seed black holes.

This paper is organized in the following way. In the second section, we briefly summarize the simulations setup and numerical methods employed. We present our results in the third section and confer our conclusions in the fourth section. 

\begin{table*}
\begin{center}
\caption{Properties of the simulated halos are listed here.}
\begin{tabular}{ccccccc}
\hline
\hline

Model	& Collapse Mass	& Spin parameter     & Redshift    & Streaming velocity at $z=100$ & Circular velocity & Sink mass \\

 & $\rm M_{\odot} $ & $\lambda$    & z  & [$\rm km/s $] & [$\rm km/s $] &  $\rm M_{\odot} $ \\ 
\hline                                                          \\
 		
A (s0)	  & $\rm 4.2 \times 10^{7}$	& 0.0309765 	&10.83  & 0 & 14.78 &  $\rm 1.7 \times 10^{5}$\\
A (s1)	  & $\rm 4.04 \times 10^{7}$	& 0.0309765 	&10.83  & 3 & 14.73 & $\rm 8.7 \times 10^{4}$\\
A (s2)	  & $\rm 5.05 \times 10^{7}$	& 0.0309765 	&10.76  & 6 & 16.35 & $\rm 2.1 \times 10^{5}$\\
A (s3)	  & $\rm 5.3 \times 10^{7}$	& 0.0309765 	&10.739 & 9 & 16.54 & $\rm 1.7 \times 10^{5}$\\ 
B (s0)    & $\rm 4.6 \times 10^{7}$     & 0.0338661     &13.51  & 0 & 15.78 & $\rm 1.7 \times 10^{5}$\\
B (s1)    & $\rm 5.2 \times 10^{7}$    & 0.0338661     &12.93  & 3 & 16.4  & $\rm 8.4 \times 10^{4}$\\
B (s2)    & $\rm 5.3 \times 10^{7}$     & 0.0338661     &13.01  & 6 & 16.54 & $\rm 7.7 \times 10^{4}$\\
B (s3)    & $\rm 5.45 \times 10^{7}$     & 0.0338661     &13.0   & 9 & 16.60 & $\rm 1.3 \times 10^{5}$\\
\hline
\end{tabular}
\label{table1}
\end{center}

\end{table*}

\section{Computational methods}

The simulations presented here are performed using the open source cosmological simulations code ENZO \citep{2004astro.ph..3044O,2013arXiv1307.2265T} which is a parallel, Eulerian, grid based, an adaptive mesh refinement code. Our simulation setup is exactly the same as described in a number of previous studies \cite{2013MNRAS.tmp.2526L,2013MNRAS.433.1607L,2013MNRAS.430..588L,2013MNRAS.432..668L}. Here, we provide a brief summary of the simulations setup and refer to the above mentioned articles for details. The simulations are started  at $z=100$ with cosmological initial conditions and two nested refinement levels in addition to the top grid are employed each with a resolution of $\rm 128^3$ cells. Our computational domain has a periodic box size of $\rm 1~Mpc/h$. We employ 15 dynamical refinement levels in the central 62 kpc region during the course of simulations and use fixed Jeans resolution of 32 cells. To simulate the evolution of dark matter, we use 5767168 particles. After reaching the maximum refinement level, we insert sink particles to follow the evolution for 10, 000 years. The details of the sinks algorithm can be found in \cite{2010ApJ...709...27W}. We use standard parameters from the WMAP 7-year data ($\Omega_{m} =0.23$, $\Omega_{b}=0.045$,$\Omega_{\lambda}=0.734$ $\rm H_{0}=71~kms^{-1}Mpc^{-1}$) with a value of $\sigma_{8} = 0.8$ \citep{2011ApJS..192...14J}. To follow the thermal evolution, we solve the rate equations of the species $\rm H$,~$\rm H^{+}$,~$\rm He$,~$\rm He^{+}$,~$\rm He^{++}$,~$\rm e^{-}$,~$\rm H^{-}$,~$\rm H_{2}$,~$\rm H_{2}^{+}$ self-consistently along with the cosmological simulations. We further presume that a strong photo-dissociating background flux of strength $\rm 10^3$ in the units of $\rm J_{21}=10^{-21}~erg~cm^{-2}~s^{-1}~Hz^{-1}~sr^{-1}$ is produced by a star forming galaxy in the vicinity of the halo with a stellar radiation temperature of $\rm 10^5$~K. Indeed, such estimates of the UV field strength are in accordance with previous studies of \cite{2008MNRAS.391.1961D} and \cite{2012MNRAS.425.2854A}. The effect of self-shielding is ignored in these calculations which may raise the strength of the critical flux even further. We use the subgrid scale turbulence model of \cite{SchmNie06c} to include unresolved turbulent fluctuations. The adaptively refined large eddy simulations (LES) approach is used to implement subgrid scale (SGS) turbulence model in AMR cosmological simulations \citep{2009ApJ...707...40M}. A detailed discussion on this topic can be found in dedicated studies \citep{SchmNie06c,2009A&A...494..127S,2013MNRAS.433.1607L}. Our approach of implementing the streaming motions is the same as in \cite{2011ApJ...736..147G}. Additional constant streaming velocities of 3, 6 and 9 km/s were added to each grid cell in the x-direction at the start of our simulations (z=100).

 
\begin{figure*}
\centering
\begin{tabular}{c c}
\begin{minipage}{4cm}
\hspace{-4cm}
\includegraphics[scale=0.4]{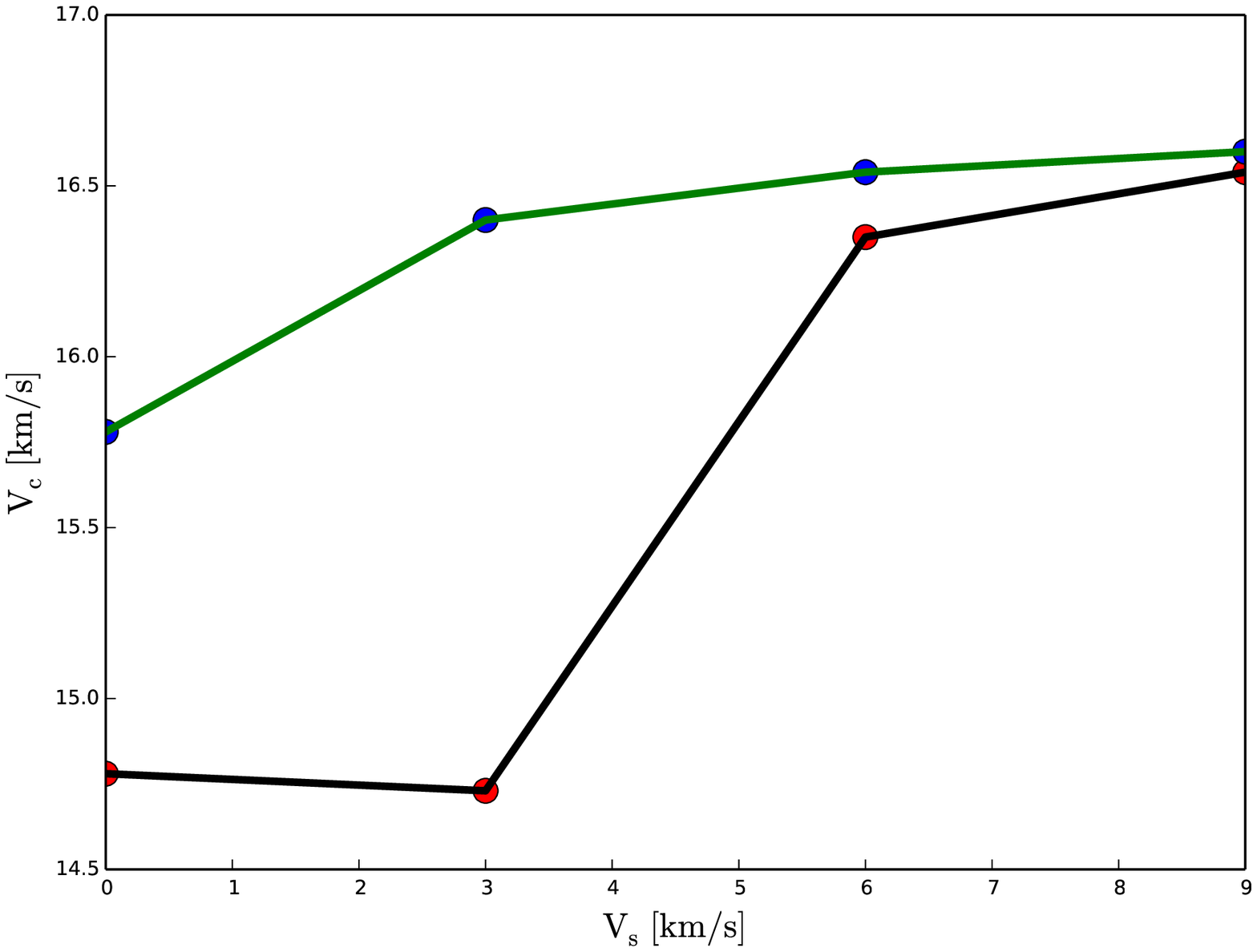}
\end{minipage} &
\begin{minipage}{4cm}
\includegraphics[scale=0.4]{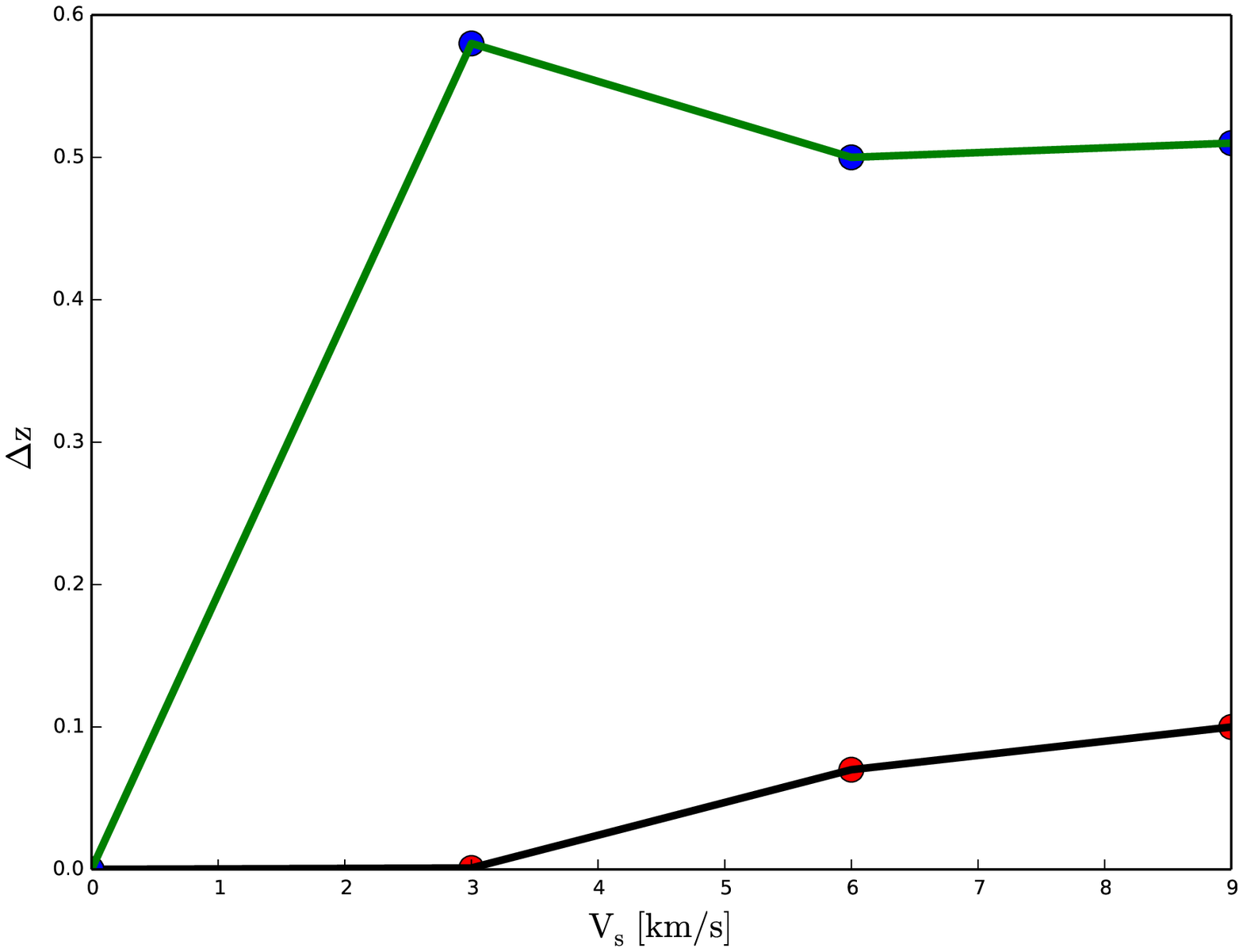}
\end{minipage} 
\end{tabular}
\caption{The circular velocities and delay in the collapse redshifts for halo A and B are shown in this figure. The circular velocity of the halo (in the left panel) and the difference is the collapse redshift are shown for the streaming velocities of 0, 3, 6 and 9 $\rm km/s$. Black line represents halo A while green line stands for halo B.}
\label{fig0}
\end{figure*}

\begin{figure*}
\hspace{-10.0cm}
\centering
\begin{tabular}{c}
\begin{minipage}{6cm}
\includegraphics[scale=0.8]{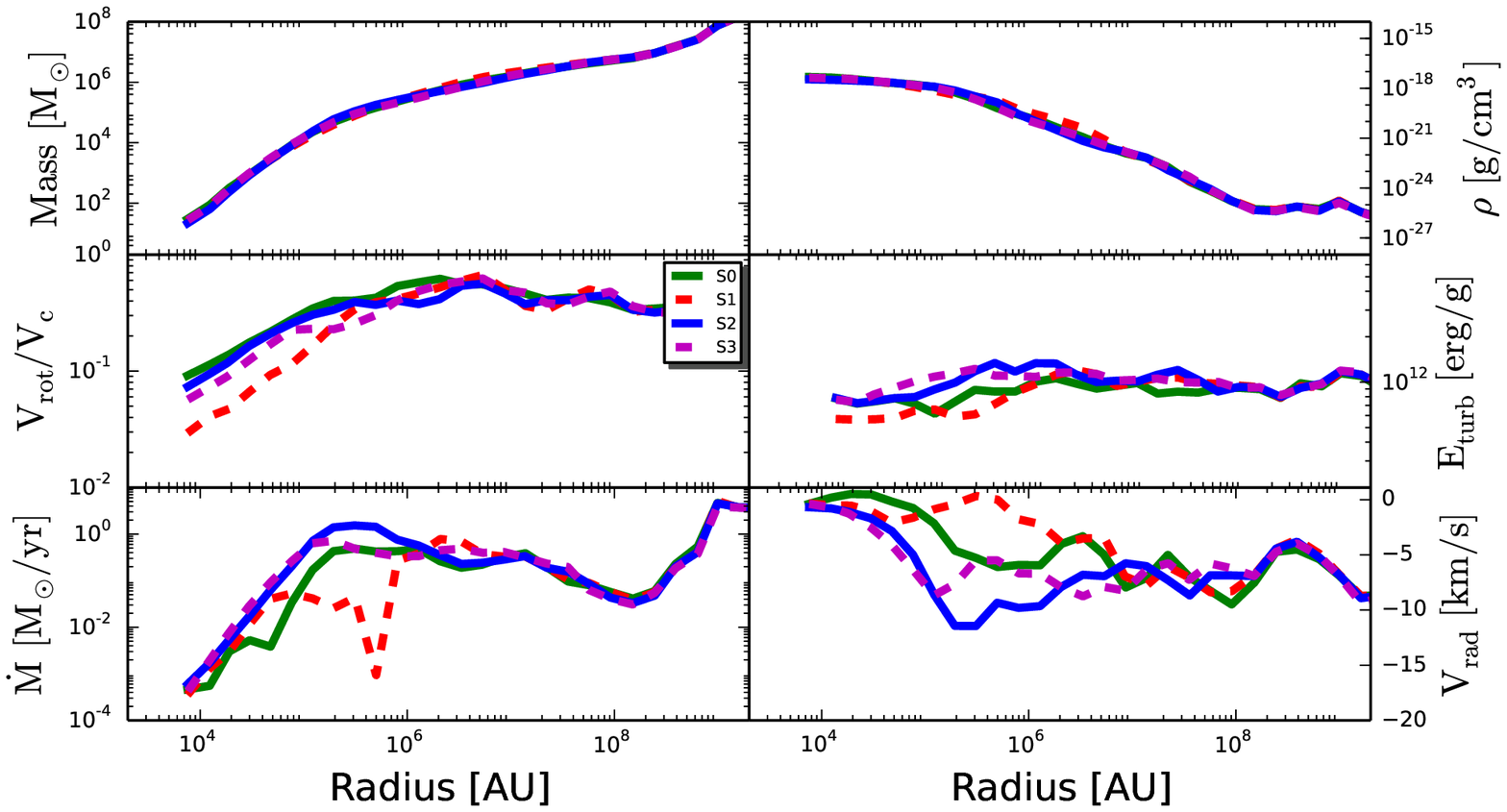}
\end{minipage}
\end{tabular}
\caption{ The properties of the halo A at its collapse redshift are shown here. The top two panel show the profiles of enclosed mass and density. Turbulent energy and the ratio of rotational to circular velocity are depicted in the middle panel. The bottom panels show the profiles of mass accretion rates and radial infall velocity. S0,S1,S2 and S3 represent the simulations with initial streaming velocities of 0, 3, 6 and 9 $\rm km/s$.}
\label{fig1}
\end{figure*}

\begin{figure*}
\hspace{-10.0cm}
\centering
\begin{tabular}{c}
\begin{minipage}{6cm}
\includegraphics[scale=0.2]{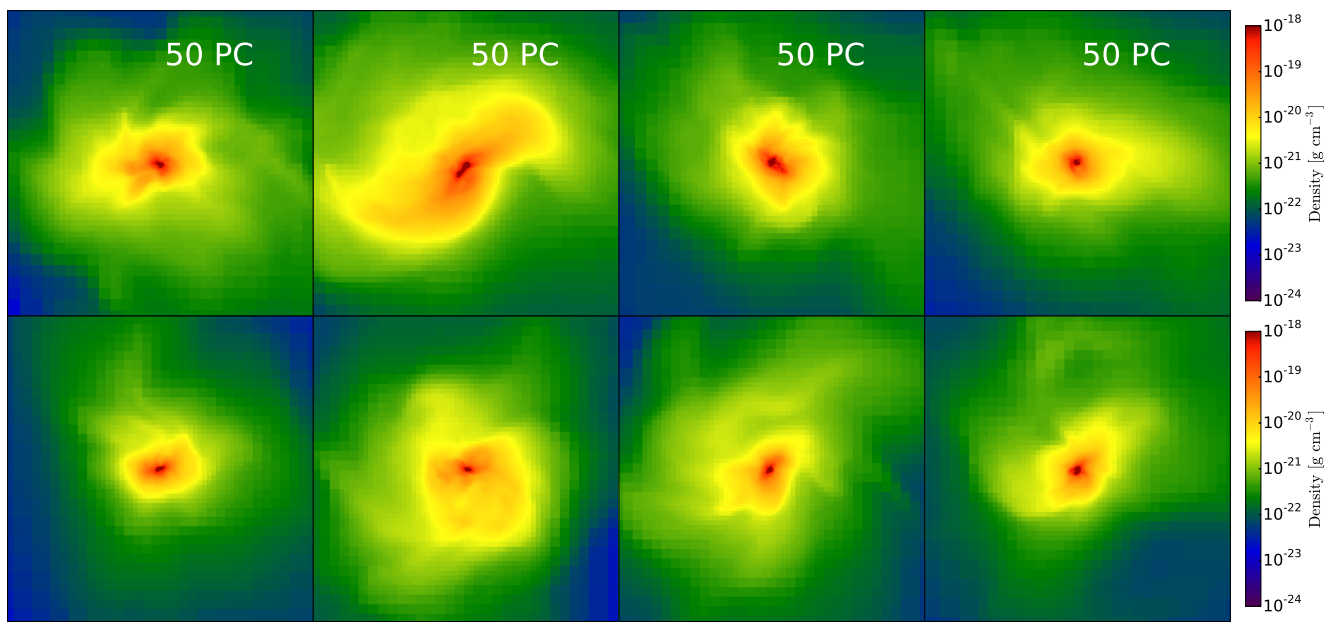}
\end{minipage}
\end{tabular}
\caption{Density projections of the simulations at their collapse redshifts are shown for the runs with various streaming velocities. Top panels show the morphology of halo A for the streaming velocities of 0-9 $\rm km/s$ (from left to right) and bottom panels show the morphologies of halo B for the streaming velocities of 0-9 $km/s$ (from left to right).}
\label{fig2}
\end{figure*}

\begin{figure*}
\hspace{-10.0cm}
\centering
\begin{tabular}{c}
\begin{minipage}{6cm}
\includegraphics[scale=0.2]{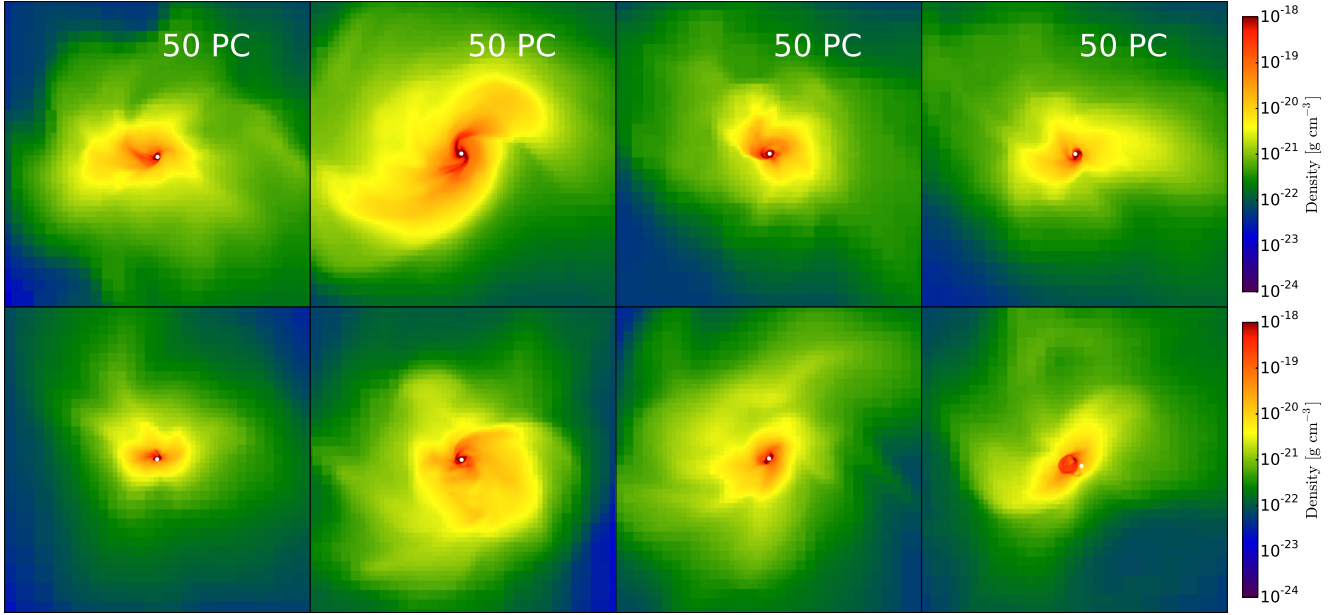}
\end{minipage}
\end{tabular}
\caption{Same as figure \ref{fig2} but at the final stage of simulations. Sink particles of the masses listed in the table \ref{table1} are overplotted on the density projections.}
\label{fig3}
\end{figure*}

\begin{figure*}
\hspace{-10.0cm}
\centering
\begin{tabular}{c}
\begin{minipage}{6cm}
\includegraphics[scale=0.8]{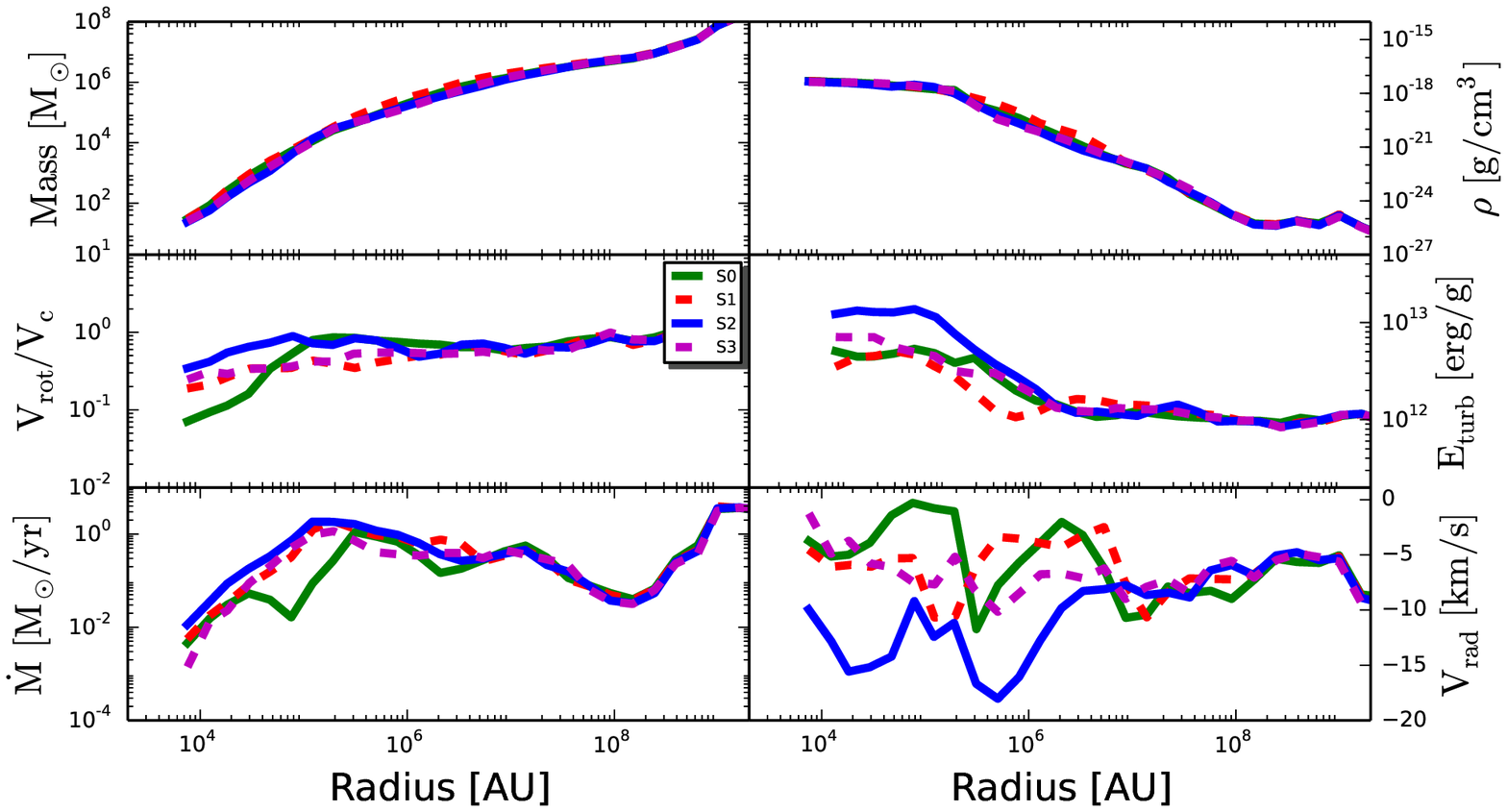}
\end{minipage}
\end{tabular}
\caption{The properties of  halo A at its final state are shown here. The top two panel show the profiles of enclosed mass and density. Turbulent energy and the ratio of rotational to circular velocity are depicted in the middle panel. The bottom panels show the profiles of mass accretion rates and radial infall velocity. S0,S1,S2 and S3 represent the simulations with initial streaming velocities of 0, 3, 6 and 9 $\rm km/s$.}
\label{fig4}
\end{figure*}

\begin{figure*}
\centering
\begin{tabular}{c c}
\begin{minipage}{4cm}
\hspace{-4cm}
\includegraphics[scale=0.4]{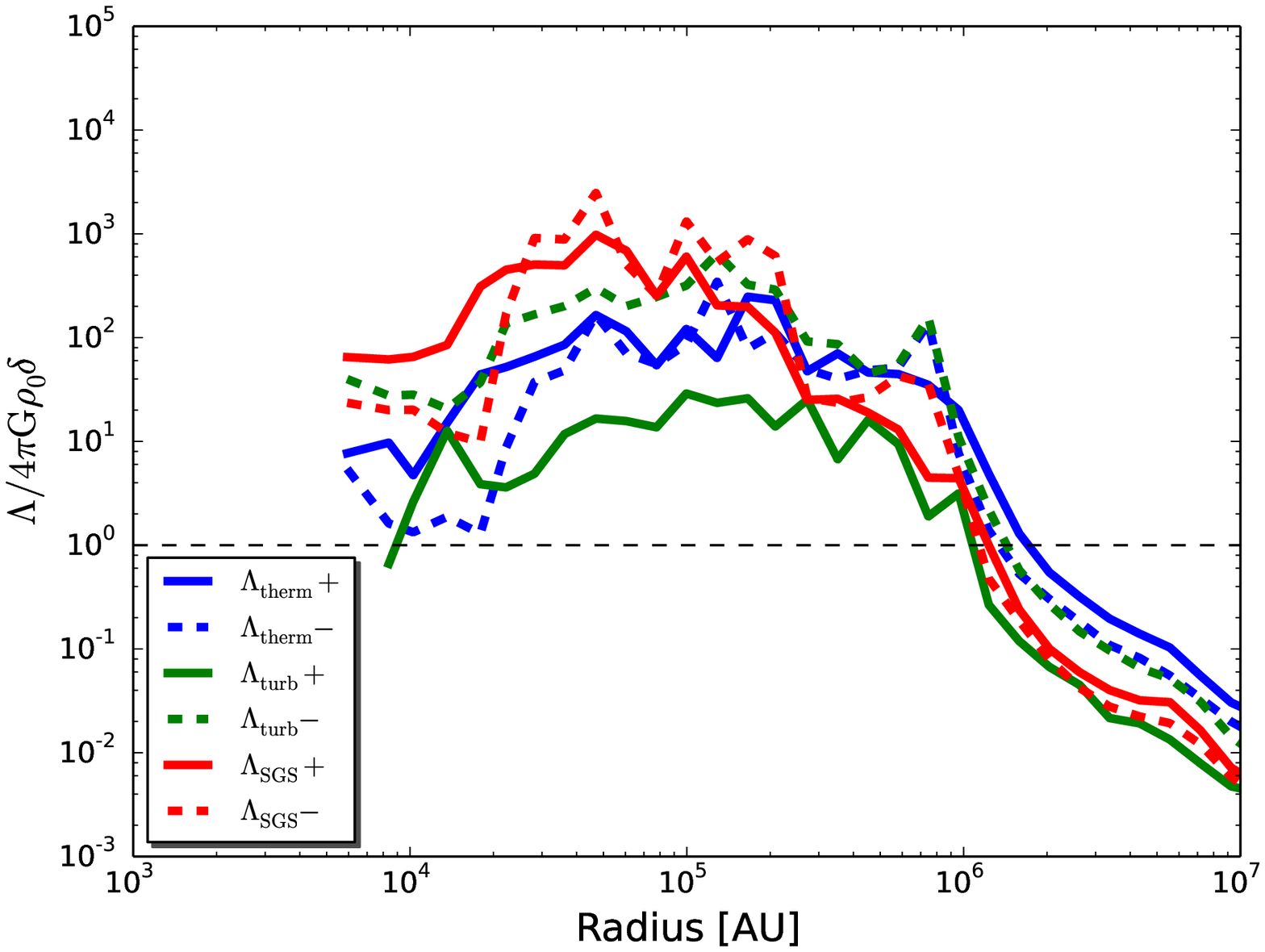}
\end{minipage} &
\begin{minipage}{4cm}
\includegraphics[scale=0.4]{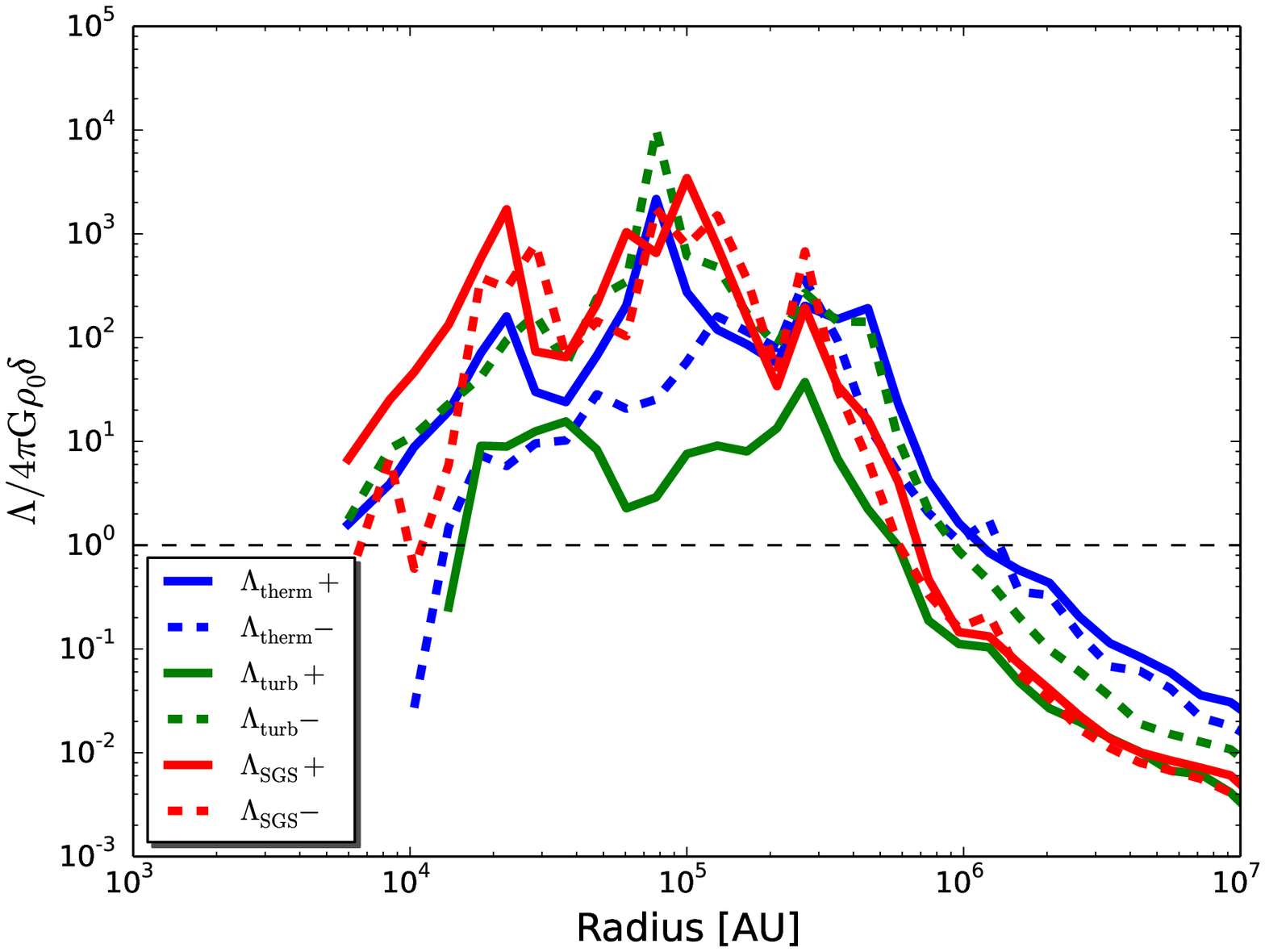}
\end{minipage} 

\end{tabular}
\caption{The contribution of support terms for halo A with streaming velocities  of zero (left) and 9 km/s (right). The local support of thermal, turbulent and SGS is shown in this figure. The solid lines show the positive support of the quantities while dashed lines represent the negative support. For the definitions of support terms see Schmidt et al. 2013.}
\label{fig5}
\end{figure*}

\section{Main Results}
In total, eight cosmological large eddy simulations are performed for initial baryonic streaming velocities of 0, 3, 6 and 9 $\rm km/s$ for two distinct halos. The properties of the simulated halos such as masses, collapse redshifts and spins are listed in table \ref{table1}. They have typical masses of a few times $\rm 10^7~M_{\odot}$ and collapse redshifts of 10.8 and 13.5, respectively. We have computed the circular velocity (also called cooling threshold velocity) which is found to be a good measure of the halo's virial temperature \citep{2012MNRAS.424.1335F} and is plotted against the strength of streaming velocity in figure \ref{fig0}. It is found that the cooling velocity increases with enhancing the magnitude of initial streaming velocity. This trend is observed for both halos and typical circular velocities are about $\rm 14.5-16.5~km/s$. Streaming motions  delay the infall of gas into the dark matter potential and enhance the halo mass which results in a higher circular velocity. We also computed the delay in halo collapse redshift which is depicted in figure \ref{fig0}. We note that the difference in the collapse redshift for zero and extreme streaming velocity cases is 0.1 and 0.5 for our simulated halos. This delay is less significant compared to the values reported for minihalos \citep{2011ApJ...736..147G}. This comes from the fact that the gravitational potential of our halos is much deeper and the halos retain their gas with a short time delay of few million years.

The properties of the halo A as a representative case at its collapse redshift are shown in figure \ref{fig1} and compared for different initial streaming velocities. The density decreases with radius and follows an $\rm R^{-2}$ behavior as expected for an isothermal collapse. The maximum density is about $\rm 10^{-18}~g/cm^{3}$. The specific turbulent energy is about $\rm 10^{12}~erg/g$ and the radial velocity is about $\rm 5-10~km/s$ which shows infall of gas into the center of the halo. The average accretion rate is about $\rm 0.1~M_{\odot}/yr$ at larger radii and decreases down to $\rm 10^{-2}~M_{\odot}/yr$ in the Jeans length. The ratio of rotational to circular velocity is 0.3. The mass profile increases linearly with radius as expected from an isothermal collapse and declines sharply within the Jeans radius, where the density profile becomes almost flat. No significant differences are observed in the above mentioned quantities with and without streaming motions. 

The state of the simulations at their collapse redshifts is represented by the density projections and is shown in figure \ref{fig2}. The maximum density is $\rm 10^{-18}~g/cm^{3}$. The morphology of the halo is slightly different for each case of streaming motions due to the different turbulence realizations in the halo. We noticed that a monolithic collapse occurs and fragmentation remains suppressed. As mentioned in the previous section, we evolved the simulations for 10,000 years by employing sink particles after reaching the maximum refinement level. Sink particles are overplotted on the density projections and are shown in figure \ref{fig3}. It is found that massive sinks of about $\rm 10^5~M_{\odot}$ are formed and their masses are listed in table \ref{table1} for the various initial streaming velocities. No clear trend is observed in the masses of sinks for the cases with streaming motions.  

We also show the central properties of halo A for a representative case at the final stage of the simulations in figure \ref{fig4}. With the passage of time, turbulent energy is increased by an order of magnitude. According to the Kolmogorov scaling, the turbulent energy should  decrease towards smaller radii for the homogenous turbulence but it is enhanced in the core of the halo for our case. This is because of the higher turbulence production rate in the center of halo mainly driven by the gravity. Such behavior  reflects the decreasing dynamical times and is noted in a number of previous studies \citep{2013MNRAS.430..588L,2013MNRAS.433.1607L,2013MNRAS.436.2989L}. The ratio of $\rm v_{rot}/v_{cir}$ is also enhanced to 0.6 which shows that the halos have a high degree of rotational support. The latter may further delay the accretion to the sink and may also slow down the collapse, see \cite{2013MNRAS.tmp.2526L}. Overall, no significant differences are observed in the halo properties after varying the streaming velocities. We further investigated the contributions of local support terms which are computed by solving the  differential equation for the rate of compressions of the gas \citep{2013MNRAS.431.3196S}. Support by thermal pressure, resolved turbulence and SGS turbulence scaled by the gravitational compression is shown for two representative cases in figure \ref{fig5}. In general, it seems that turbulence support becomes important in the core of the halo and extends out to radii of 10 pc. Particularly, support of SGS turbulence  becomes important in the core of the halo and yields relevant contributions to the support against gravity. Overall, no qualitative difference is found in the support terms for the cases with and without streaming motions.

\section{Discussion}

In all, we have performed eight cosmological large eddy simulations for two distinct halos to study the impact of baryonic streaming motions on the formation of supermassive black holes via the direct collapse scenario. To accomplish this goal, we introduced the baryonic streaming velocities of 3, 6 and 9 $\rm km/s$ in the x-direction at $z=100$, and compared the results with and without streaming motions. We added 15 dynamical refinement levels and a fixed Jeans resolution of 32 cells during the entire course of simulations. We further employed sink particles to follow the accretion on them for 10,000 years after reaching the maximum refinement level.

Our findings show that the threshold circular velocity which defines the minimum cooling mass of the halo is about 14-16 $\rm km/s$ and gets enhanced in the presence of large streaming velocities. The overall increase in the circular velocity is 1-2 $\rm km/s$ and less significant compared to the minihalos. This is due to the fact that the halos in our simulations have masses of a few times $\rm 10^{7}~M_{\odot}$, their gravitational potentials are much deeper and can retain sufficient gas even in the presence of baryonic streaming motions. We also noticed that such an increase in the circular velocity delays the collapse for a few million years. Again, in the minihalos delay was $\Delta z = 4$ and in our case it is about $\Delta z = 0.5$. 

Streaming velocities decay as $1+z$, and their impact may become more significant if such halos are formed at redshift around 30. In this case, the energy ratio  of streaming to circular velocity (i.e., $v_{s}^2/v_{c}^2$) may increase by a factor of 4, making their impact potentially more significant. Such halos would thus correspond to high-sigma peaks, but the impact is likely still reduced compared to minihalos.

No significant differences are found in the general properties of the halos such as density, radial velocity, turbulent energy and accretion rates with and without baryonic motions. Sink particles have typical masses of $\rm 10^5~M_{\odot}$ and are potential candidates for the formation of supermassive stars as an intermediate stage to supermassive black holes \citep{2013A&A...558A..59S,2012ApJ...756...93H,2011MNRAS.414.2751B,2013ApJ...771..116J,2013arXiv1308.4457H}. The observational imprint of such stars can be probed with upcoming telescopes such as JWST \footnote{http://www.jwst.nasa.gov/} and ATHENA+\footnote{http://athena2.irap.omp.eu/} \citep{2013ApJ...777...99W,2014ApJ...781..106W}. No systematic trend is observed in the masses of sinks for streaming motions. We have explored here even the extreme cases with streaming velocities of 6 and 9 $\rm km/s$ and their impact on the halo properties and sink masses is negligible. 

In a recent study \cite{2013arXiv1310.0859T} have proposed that extreme streaming velocities may facilitate the direct collapse scenario by enhancing the critical mass for the collapse to commence, suppressing the formation of stars and consequently avoiding the metal enrichment. This study further suggests that under such conditions there is no need of background UV flux to quench $\rm H_{2}$ formation as the later can be dissociated by the collisional shocks at higher densities. This scenario requires additional mechanisms to suppress the $\rm H_{2}$ cooling at higher densities. In a very recent work \cite{2014arXiv1401.5803F} have explored the feasibility of collisional dissociation of molecular hydrogen in the presence of shocks and found that it is very difficult to avoid the formation of $\rm H_{2}$ without background UV flux. Therefore, the need of a strong background UV flux seems necessary for the formation of massive black holes. Nevertheless, it is desirable to extend investigation of baryonic streaming velocities to massive halos where $\rm H_{2}$ cooling is relevant in future studies.

%


\section*{Acknowledgments}
The simulations described in this work were performed using the Enzo code, developed by the Laboratory for Computational Astrophysics at the University of California in San Diego (http://lca.ucsd.edu). We thank Wolfram Schmidt for useful discussions on the topic. We acknowledge research funding by Deutsche Forschungsgemeinschaft (DFG) under grant SFB $\rm 963/1$ (projects A12, A15) and computing time from HLRN under project nip00029. DRGS thanks the DFG for funding via the Schwerpunktprogram SPP 1573 ``Physics of the Interstellar Medium'' (grant SCHL $\rm 1964/1-1$). The simulation results are analyzed using the visualization toolkit for astrophysical data YT \citep{2011ApJS..192....9T}.

\bibliography{blackholes.bib}

\end{document}